\documentclass[aps,prl,reprint,superscriptaddress,nofootinbib]{revtex4-2}

\usepackage{graphicx}
\usepackage{amsmath}

\usepackage{amssymb}

\usepackage{overpic}
\usepackage[colorlinks=true,allcolors=blue]{hyperref}

\usepackage[colorlinks=true,allcolors=blue]{hyperref}

\makeatletter
\let\frontmatter@footnote@produce
    \frontmatter@footnote@produce@footnote
\makeatother

\begin{document}

\title{Self-Synchronized Terahertz and X-Ray Free-Electron Lasers from a Single Pre-Bunched Electron Beam}

\author{Yin Kang}
\affiliation{Shanghai Advanced Research Institute, Chinese Academy of Sciences, Shanghai 201210, China}

\author{Kaiqing Zhang}
\email[Corresponding author: ]{zhangkq@sari.ac.cn}
\affiliation{Shanghai Advanced Research Institute, Chinese Academy of Sciences, Shanghai 201210, China}
\affiliation{University of Chinese Academy of Sciences, Beijing, China} 

\author{Zhen Wang}
\affiliation{Shanghai Advanced Research Institute, Chinese Academy of Sciences, Shanghai 201210, China}
\affiliation{University of Chinese Academy of Sciences, Beijing, China} 

\author{Cheng Yu}
\affiliation{Shanghai Advanced Research Institute, Chinese Academy of Sciences, Shanghai 201210, China}

\author{Zhangfeng Gao}
\affiliation{Shanghai Advanced Research Institute, Chinese Academy of Sciences, Shanghai 201210, China}

\author{Wencai Cheng}
\affiliation{Shanghai Advanced Research Institute, Chinese Academy of Sciences, Shanghai 201210, China}

\author{Hang Luo}
\affiliation{Shanghai Advanced Research Institute, Chinese Academy of Sciences, Shanghai 201210, China}

\author{Yue Wang}
\affiliation{Shanghai Advanced Research Institute, Chinese Academy of Sciences, Shanghai 201210, China}

\author{Hanghua Xu}
\affiliation{Shanghai Advanced Research Institute, Chinese Academy of Sciences, Shanghai 201210, China}

\author{Xiaoqing Liu}
\affiliation{Shanghai Advanced Research Institute, Chinese Academy of Sciences, Shanghai 201210, China}

\author{Jinguo Wang}
\affiliation{Shanghai Advanced Research Institute, Chinese Academy of Sciences, Shanghai 201210, China}

\author{Huan Zhao}
\affiliation{Shanghai Advanced Research Institute, Chinese Academy of Sciences, Shanghai 201210, China}

\author{Yanyan Zhu}
\affiliation{Shanghai Advanced Research Institute, Chinese Academy of Sciences, Shanghai 201210, China}

\author{Yongmei Wen}
\affiliation{Shanghai Advanced Research Institute, Chinese Academy of Sciences, Shanghai 201210, China}

\author{Fei Gao}
\affiliation{Shanghai Advanced Research Institute, Chinese Academy of Sciences, Shanghai 201210, China}

\author{Yangyang Lei}
\affiliation{Shanghai Advanced Research Institute, Chinese Academy of Sciences, Shanghai 201210, China}

\author{Chengcheng Xiao}
\affiliation{Shanghai Advanced Research Institute, Chinese Academy of Sciences, Shanghai 201210, China}

\author{Liping Sun}
\affiliation{Shanghai Advanced Research Institute, Chinese Academy of Sciences, Shanghai 201210, China}

\author{Yongfang Liu}
\affiliation{Shanghai Advanced Research Institute, Chinese Academy of Sciences, Shanghai 201210, China}

\author{Jiaqiang Xu}
\affiliation{Shanghai Advanced Research Institute, Chinese Academy of Sciences, Shanghai 201210, China}

\author{Weiyi Yin}
\affiliation{Shanghai Advanced Research Institute, Chinese Academy of Sciences, Shanghai 201210, China}

\author{Xingtao Wang}
\affiliation{Shanghai Advanced Research Institute, Chinese Academy of Sciences, Shanghai 201210, China}

\author{Taihe Lan}
\affiliation{Shanghai Advanced Research Institute, Chinese Academy of Sciences, Shanghai 201210, China}

\author{Zheng Qi}
\affiliation{Shanghai Advanced Research Institute, Chinese Academy of Sciences, Shanghai 201210, China}

\author{Tao Liu}
\affiliation{Shanghai Advanced Research Institute, Chinese Academy of Sciences, Shanghai 201210, China}

\author{Zhi Guo}
\affiliation{Shanghai Advanced Research Institute, Chinese Academy of Sciences, Shanghai 201210, China}
\affiliation{University of Chinese Academy of Sciences, Beijing, China} 

\author{Bin Li}
\affiliation{Shanghai Advanced Research Institute, Chinese Academy of Sciences, Shanghai 201210, China}
\affiliation{University of Chinese Academy of Sciences, Beijing, China} 

\author{Chao Feng}
\email[Corresponding author: ]{fengc@sari.ac.cn}
\affiliation{Shanghai Advanced Research Institute, Chinese Academy of Sciences, Shanghai 201210, China}
\affiliation{University of Chinese Academy of Sciences, Beijing, China} 

\author{Bo Liu}
\affiliation{Shanghai Advanced Research Institute, Chinese Academy of Sciences, Shanghai 201210, China}
\affiliation{University of Chinese Academy of Sciences, Beijing, China} 

\author{Zhentang Zhao}
\affiliation{Shanghai Advanced Research Institute, Chinese Academy of Sciences, Shanghai 201210, China}
\affiliation{University of Chinese Academy of Sciences, Beijing, China} 

\begin{abstract}
Ultrafast pump-probe spectroscopy combining intense terahertz (THz) and X-ray pulses is a critical tool for investigating complex structural and electronic dynamics in materials. However, current setups combining THz sources and X-ray free-electron lasers (FELs) often suffer from high system complexity, inherent timing jitter, or limited THz pulse properties. Here, we experimentally demonstrate the generation of intrinsically synchronized, strong-field, narrow-band THz and X-ray FELs from a single pre-bunched electron beam. Sequentially passing the beam through X-ray and THz amplifiers reveals a highly synergistic process: the initial periodic THz density modulation notably boosts the X-ray FEL pulse energy, while robustly surviving the intense X-ray emission to drive high-power, narrow-band THz radiation. Originating from the same electron bunch, the two pulses inherently maintain a precise, constant time delay. This jitter-free scheme establishes a highly reliable platform tailored for both X-ray-pump/THz-probe and THz-pump/X-ray-probe experiments. 
\end{abstract}

\maketitle

\begingroup
\renewcommand{\thefootnote}{\fnsymbol{footnote}}

    

\endgroup

Ultrafast pump-probe spectroscopy combining terahertz (THz) and X-ray pulses is a  powerful tool for investigating complex dynamics in condensed matter physics and materials science \cite{thzx_kampfrath2013_resonant_nonresonant_control,thzx_zhang2017_extreme_terahertz_science,thzx_salen2019_matter_manipulation,thzx_riepp2024_coherent_magnetization,thzx_johnson2025_xrayview}. Spanning $0.1$ to $30\text{ THz}$, MV/cm-scale strong-field THz radiation directly resonates with fundamental collective excitations, such as lattice vibrations \cite{thzx_basini2024_multiferroicity,thzx_fechner2024_lattice,thzx_kozina2019_upconversion} and spin precessions \cite{thzx_kubacka2014_electromagnon,thzx_mashkovich2021_spinlattice,thzx_zhang2024_magnon,thzx_ilyas2024_magnetization}, selectively driving materials into far-from-equilibrium states \cite{thzx_li2019_ferroelectricity,thzx_yamakawa2021_chargeorder,thzx_shi2023_mote2,thzx_delatorre2021_nonthermal,thzx_venanzi2024_trions,thzx_guo2026_plasmonic_time_crystal,thzx_ueda2023_spinlattice}. To capture these fast processes, X-ray free-electron lasers (FELs) provide simultaneous controllable temporal and nanometer spatial resolution, mapping the ensuing structural and electronic responses at the atomic level \cite{thzx_li2021_polarvortices,thzx_orenstein2025_polarizationwaves,thzx_wang2025_skyrons,thzx_johnson2023_vo2,thzx_pellegrini2016_xfelphysics}. In a complementary way, an X-ray-pump and THz-probe setup can track transient exciton diffusion and carrier dynamics after core-level absorption \cite{thzx_zapolnova2020_plasmaswitch,thzx_chen2021_ultrafast_conductivity,thzx_kubota2025_xraythz}. Realizing the full experimental potential of these techniques, however, strictly requires strong-field, frequency-tunable THz and X-ray pulses with rigorous timing synchronization.

A direct way to achieve this THz and X-ray synchronization is to generate both pulses from the same relativistic electron beam. Major high-gain FEL facilities (such as FERMI, LCLS, and FLASH) have generated THz radiation using coherent transition radiation (CTR) or undulator emission from highly compressed bunches \cite{thzx_wu2013_ctr,thzx_dimitri2018_terafermi_ctr,thzx_zapolnova2018_flash_doubler,thzx_pan2019_flash_diagnostics,thzx_zhang2020_lcls2thz}. However, these methods rely on coherent emission rather than high-gain amplification. As a result, the output is either broadband or weak (typically $\textless 1\text{ MV/cm}$ with filters), and the field strength drops quickly at higher frequencies \cite{thzx_riepp2024_coherent_magnetization}. In addition, strong bunch compression makes it hard to control the duration and profile of the X-ray FEL pulse. To achieve stronger THz fields, some facilities use external laser-driven THz sources, such as optical rectification or laser plasmas \cite{thzx_kubacka2014_electromagnon,thzx_kozina2019_upconversion,thzx_li2021_polarvortices,thzx_ueda2023_spinlattice,thzx_orenstein2025_polarizationwaves,thzx_wang2025_skyrons}. These external sources require complex optical systems and also introduce extra timing jitter, which reduces the temporal resolution of pump-probe experiments. Therefore, generating intense, tunable, and naturally synchronized THz and X-ray pulses remains a central challenge.

\begin{figure*}[htbp]
    \centering
    \includegraphics[width=0.85\textwidth]{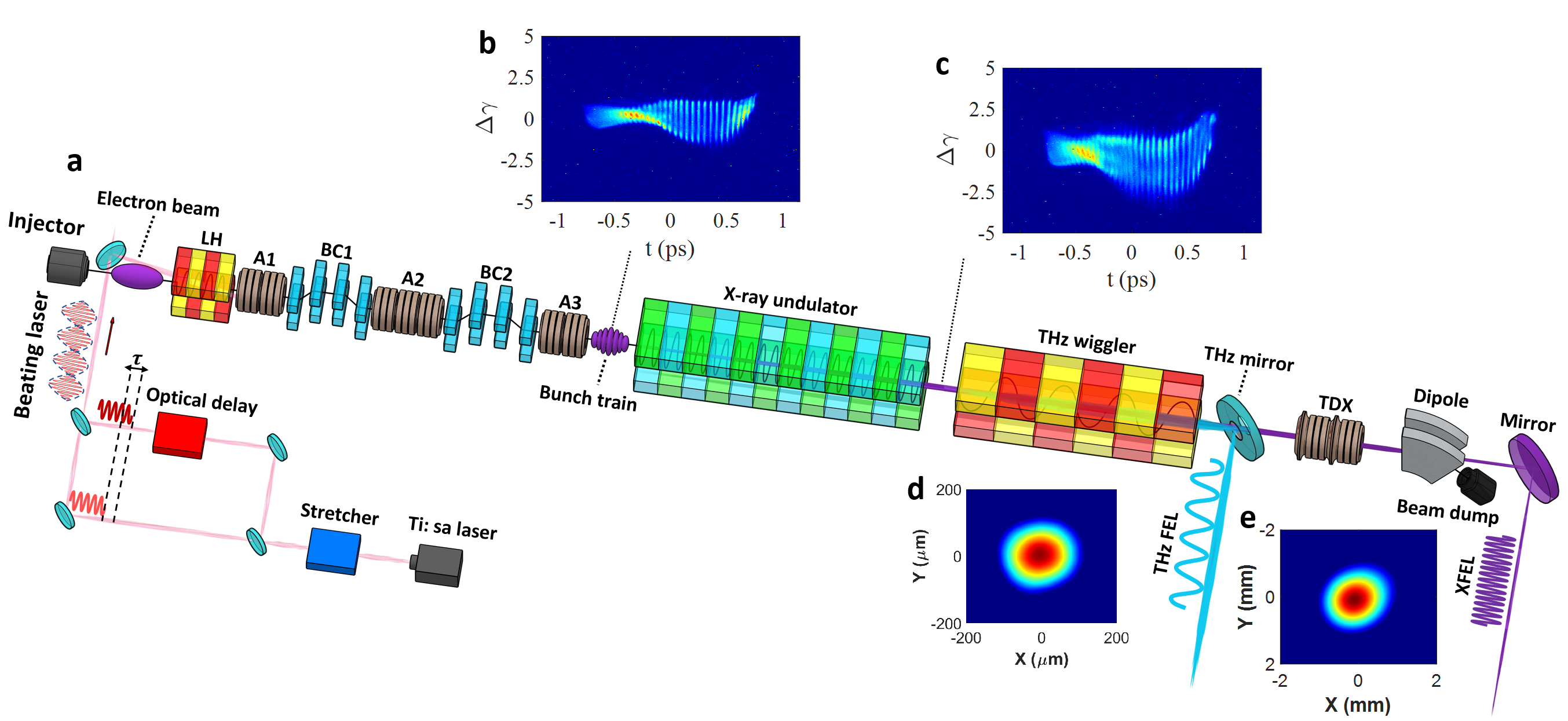}  
    \caption{Schematic of the experimental setup based on Shanghai soft X-ray free-electron laser facility. (a) An electron beam interacts with a beating laser in the laser heater (LH) to obtain energy modulation, which is converted to a density modulation and frequency-upshifted through the bunch compressors (BCs). The pre-bunched beam passes sequentially through the X-ray undulator and THz wiggler to generate synchronized X-ray FEL and THz radiation. (b), (c) Measured longitudinal phase spaces before and after the X-ray undulator. (d), (e) Measured transverse profiles of the THz and X-ray radiation spots.}
    \label{fig:1}
\end{figure*}

Recently, we demonstrated a strong-field THz source (up to 60 MV/cm) based on longitudinal beam shaping, which provides narrowband and continuously tunable radiation \cite{thzx_kang2026_continuous}. However, whether such a shaped beam can simultaneously drive an X-ray FEL remained a challenging question. Physically, since THz and X-ray wavelengths differ by more than three orders of magnitude, their respective requirements on the electron-beam phase space can be decoupled \cite{thzx_pan2019_flash_diagnostics,thzx_kang2023_xfelthz}. However, in practice, the THz density modulation can degrade the local beam quality by increasing the slice energy spread, which usually suppresses the high-gain X-ray FEL process. Furthermore, the intense X-ray radiation itself induces energy diffusion that may degrade the THz bunching structure. The critical challenge is to balance these two scalable processes by preserving the local beam quality required for X-ray amplification while keeping a robust density modulation for downstream THz-FEL amplification.

In this Letter, we present the first demonstration of self-synchronized THz and X-ray FELs at the Shanghai soft X-ray free-electron laser facility (SXFEL), by passing a pre-bunched electron beam (at THz frequency) sequentially through an X-ray undulator and a subsequent THz wiggler. By optimizing the initial phase-space modulation, the THz modulation shapes the X-ray radiation into distinct pulse trains, leading to a substantial enhancement of the overall X-ray FEL pulse energy due to the enhanced peak current of the electron beam. Concurrently, the initial THz density modulation robustly survives the intense X-ray emission process, successfully driving high-power THz radiation in the subsequent wiggler. This jitter-free setup establishes a reliable platform well suited for both strong-field THz-pump/X-ray-probe and X-ray-pump/THz-probe experiments \cite{hafez2016intense,thzx_gensch2008_flash_undulator,thzx_fruhling2009_xray_streak_camera,PhysRevLett.122.073001}.

Fig.~\ref{fig:1}a illustrates the facility layout and experimental schematic, comprising an injector, a laser heater (LH), a main accelerator with two-stage bunch compressors (BCs), ten X-ray undulators, and a downstream THz wiggler. The physical process begins in the LH, where a $400\text{-pC}$, $115\text{-MeV}$ electron beam (energy spread $80\text{ keV}$, normalized transverse emittance $1~\mathrm{mm}\cdot\mathrm{mrad}$, bunch length $\sim 14.3\text{ ps}$) co-propagates with a frequency-beating optical field generated via the optical heterodyning of an $800\text{-nm}$ Ti:sapphire laser. This interaction purposely imprints a relatively small initial energy modulation onto the beam, with a continuously tunable fundamental frequency of $f_0 = \mu \tau / 2\pi$, where $\mu$ is the linear chirp rate and $\tau$ is the adjustable relative time delay of the beating lasers \cite{thzx_weling1996_narrowband,thzx_dunning2012_periodic}. Crucially, this small energy modulation governs the entire beam evolution by playing a dual role through the microbunching instability (MBI) effects. On the one hand, this coherent seed effectively suppresses the chaotic, noise-driven MBI that would otherwise degrade the global beam quality. On the other hand, as the beam accelerates to $1.04\text{ GeV}$ and passes through the BCs, the MBI mechanism itself drives a continuous, self-reinforcing conversion between energy and density modulations \cite{thzx_dunning2012_periodic,thzx_musumeci2011_spacecharge,thzx_saldin2004_microbunching}. Through this controlled MBI process, the initially small energy modulation is strongly amplified and transformed into a highly regular, periodic density modulation, while its frequency is up-shifted to $C f_0$ ($C \approx 10$ being the compression factor of the two BCs).

The resulting longitudinal phase space, measured by an X-band transverse deflecting cavity (TDX), shows the structure of THz pulse trains with enhanced local peak currents (Fig.~\ref{fig:1}b). This provides a pre-bunched beam with a continuously tunable frequency from $2\text{--}30\text{ THz}$ \cite{thzx_brynes2020_mbi_modulated_laser,thzx_kang2026_continuous}. The beam sequentially passes through 10 $16\text{-mm}$-period X-ray undulator segments (each 4 m long) to drive FEL emission at $4.1\text{ nm}$, and then a $5\text{-m}$-long electromagnetic wiggler (period $280\text{ mm}$) to generate THz radiation. Because the THz and X-ray wavelengths differ by orders of magnitude, they have different tolerances to the beam energy spread. The phase space measured after the X-ray undulator (Fig.~\ref{fig:1}c) confirms that the THz microbunching structure survives the X-ray emission process, allowing the beam to drive THz radiation downstream. Afterward, an apertured mirror is used downstream to separate the co-propagating THz and X-ray pulses.

\begin{figure}[htbp]
    \centering
    \includegraphics[width=0.4\textwidth]{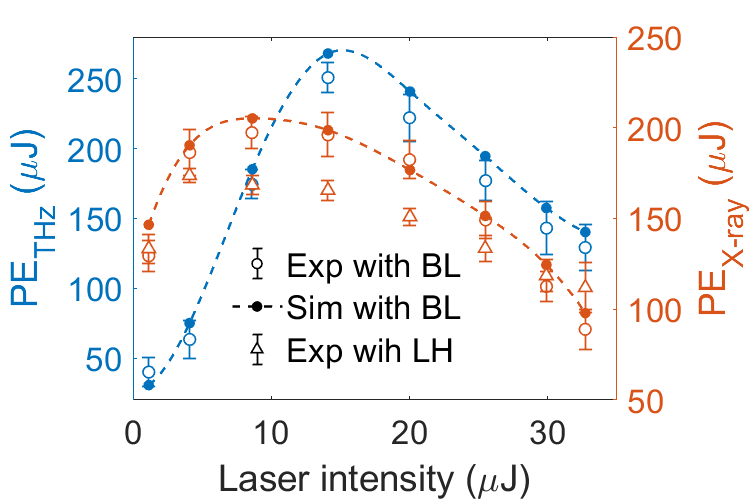}  
    \caption{Radiation pulse energies (PE) versus beating laser (BL) intensity. Measured (Exp) and simulated (Sim) PE of THz radiation (left axis) and X-ray FEL (right axis) as a function of the BL intensity under $19\text{-THz}$ modulation. The X-ray PE with a conventional LH (unmodulated single laser pulse) are shown for comparison.}
    \label{fig:2}
\end{figure}

\begin{figure*}[htbp]
    \centering
    \begin{overpic}[width=0.36\linewidth, height=5.2cm]{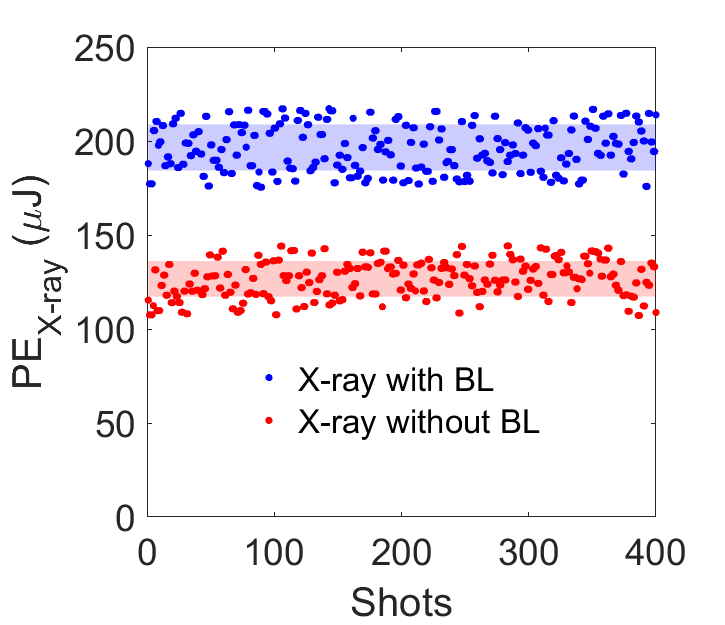}
        \put(0.5, 78){\textbf{(a)}} 
    \end{overpic}
    \hspace{0.5cm} 
    \begin{overpic}[width=0.32\linewidth, height=5cm]{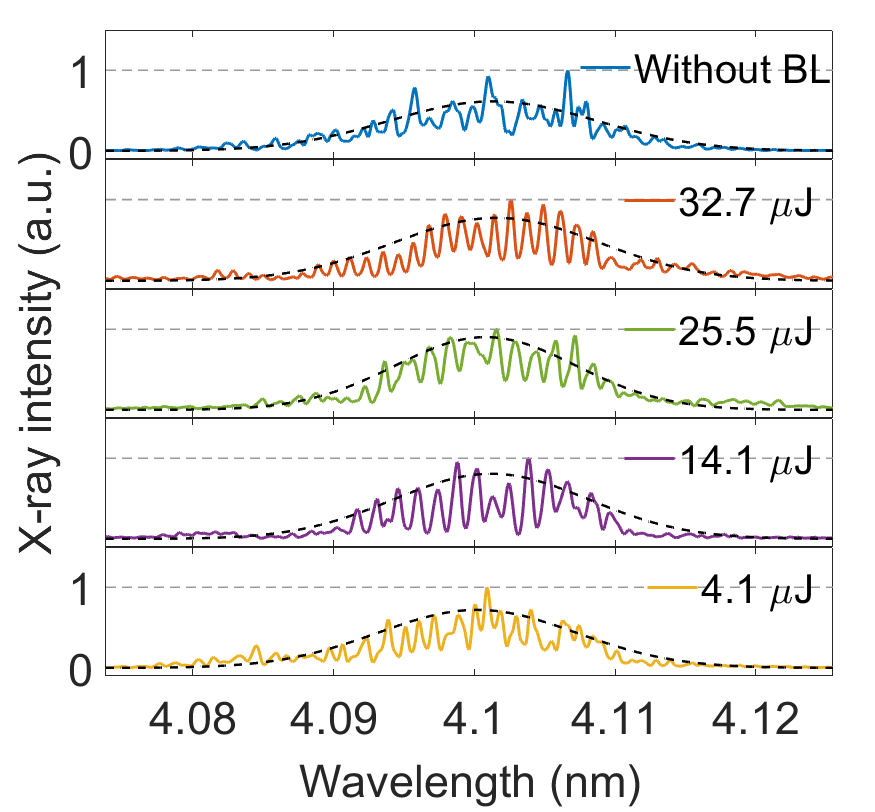}
        \put(-0.0, 87){\textbf{(b)}}
    \end{overpic}
    \caption{Measured X-ray FEL performances with and without THz modulation (beating laser, BL). (a) Measured X-ray pulse energies (PE) over 400 consecutive shots for the THz-modulated and unmodulated cases. (b) Measured X-ray spectra under THz modulation at different beating laser pulse energies, benchmarked against the unmodulated case.}
    \label{fig:3}
\end{figure*}

Before the X-ray undulator, the MBI-amplified density modulation enhances the local beam current $I$ by a factor of $B$, giving $I = B I_0$. Based on longitudinal phase-space conservation, the local relative energy spread degrades proportionally to $\sigma_\gamma / \gamma = B \sigma_{\gamma0} / \gamma$ \cite{thzx_zholents2005_esase,thzx_saldin2000_felbook}. In the 1D limit, the Pierce parameter scales as $\rho_{\text{1D}} \propto I^{1/3}$, which reduces the 1D gain length to $L_{\text{1D}} = B^{-1/3}L_{\text{1D0}}$. Including 3D effects via the Ming Xie formalism \cite{thzx_xie1995_optimization}, the 3D gain length scales as $L_{\text{3D}} = L_{\text{1D0}} B^{-1/3} (1 + 4.90 B^{1.31} \eta_{\gamma0}^{1.96}+{\Lambda}_{(d+\epsilon)}))$, where $\eta_{\gamma0} = \frac{1}{3} \frac{\sigma_{\gamma0}/\gamma}{\rho_{\text{1D}}}$ is the initial dimensionless energy-spread parameter and ${\Lambda}_{(d+\epsilon)}$  is the 3D correction factor of diffraction and emittance parameters. The saturated peak power is then given by $P_s \propto \rho_{\text{1D}} P_{\text{beam}} (L_{\text{1D0}} / L_{\text{3D}})^2$, where $P_{\text{beam}}$ is the electron beam power \cite{thzx_bonifacio1984_highgain,thzx_saldin2000_felbook}. This scaling indicates a direct trade-off for X-ray generation: the MBI-driven current enhancement increases the FEL gain, while the induced energy spread suppresses it. Because the accelerator parameters are fixed, the MBI enhancement of the THz bunching depends largely on the initial energy modulation. Therefore, optimal initial beating laser intensities exist to maximize the X-ray and THz pulse energies.  

To investigate this, we experimentally scanned the beating laser intensity using a $19\text{ THz}$ modulation as a representative case. Fig.~\ref{fig:2} shows the measured X-ray and THz pulse energies, monitored by a calibrated photodiode and a Golay cell, respectively. The X-ray energy increases at moderate laser intensities below $\sim 1.1\ \mu\text{J}$, and then decreases beyond $29\ \mu\text{J}$ when the energy spread effect becomes dominant. By contrast, the THz pulse energy reaches its maximum at a laser intensity of $\sim 14\ \mu\text{J}$. To further verify these observations, start-to-end simulations \cite{thzx_kang2026_continuous} were performed using ASTRA \cite{thzx_flottmann2017_astra} for the injector, ELEGANT \cite{thzx_borland2000_elegant} for the linac, and GENESIS \cite{thzx_reiche1999_genesis} for the FEL process. The simulated results, plotted alongside the measurements in Fig.~\ref{fig:2}, agree well with the experimental data, confirming the MBI-driven beam dynamics and the resulting radiation performance.

\begin{figure*}[htbp]
     \centering
    \begin{overpic}[width=0.4\linewidth, height=5.0cm]{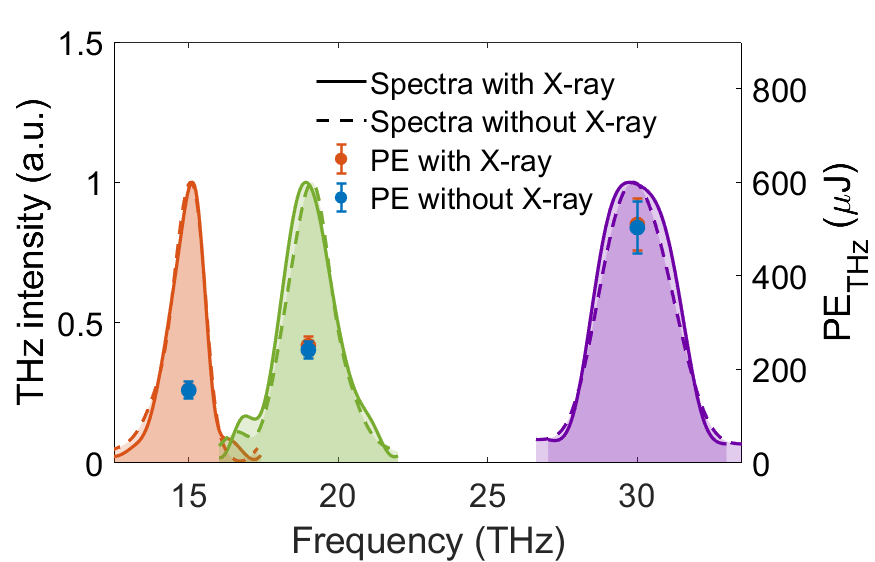}
        \put(0.4, 68){\textbf{(a)}} 
    \end{overpic}
    \hspace{0.5cm} 
    \begin{overpic}[width=0.36\linewidth, height=5cm]{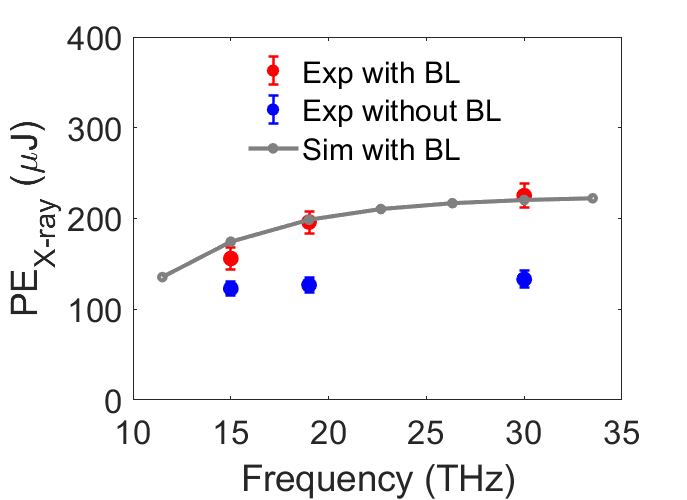}
        \put(-0.0, 75){\textbf{(b)}}
    \end{overpic}
    \caption{Radiation performances across different THz modulation frequencies. (a) Measured THz spectra and pulse energies (PE) at selected modulation frequencies ($15$, $19$ and $30\text{ THz}$) with and without upstream X-ray lasing. (b) Measured (Exp) and simulated (Sim) X-ray pulse energies versus modulation frequency, benchmarked against the unmodulated baseline.}
    \label{fig:4}
\end{figure*}

Guided by the previously established MBI-driven competition dynamics, the optimized beating laser energy was chosen to be $14\ \mu\text{J}$ to safely sustain the FEL amplification (Fig.~\ref{fig:2}). To systematically evaluate the performance, X-ray pulse energies were recorded over 400 consecutive shots. As illustrated in Fig.~\ref{fig:3}a, the beam modulated by the optimized THz envelope drives a notably enhanced mean X-ray pulse energy of $197\ \mu\text{J}$ and  a root mean square (RMS) relative fluctuation of $6.23\ \%$, compared with the $127\ \mu\text{J}$ and $7.5\ \%$ for the unmodulated beam. As theoretically predicted by the 3D FEL scaling laws, this confirms that the periodic current enhancement boosts the FEL output power, effectively overcoming the detrimental effect of the increased local energy spread.

Furthermore, the local current enhancement shapes the longitudinal profile of the X-ray emission into a train of sub-pulses \cite{thzx_marinelli2016_opticalshaping,thzx_duris2021_pulsetrains}. The temporal electric field $E(t)$ can be modeled as a coherent superposition of $N$ equally spaced pulses:
\begin{equation}
E(t) = \sum_{j=1}^{N} E_{0j} \exp\left( -\frac{(t-t_{0j})^2}{4\sigma_t^2} - i\omega_0(t-t_{0j}) + i\phi_j \right),
\end{equation}
where $E_{0j}$, $t_{0j}$, and $\phi_j$ are the amplitude, temporal peak position, and phase of the $j$-th sub-pulse, $\sigma_t$ is the pulse duration, and $\omega_0$ is the central angular frequency of X-ray. Interference among these sub-pulses creates a quasi-periodic spectral fringe pattern \cite{thzx_marinelli2016_opticalshaping,thzx_duris2021_pulsetrains,thzx_hu2025_modelockedcomb}. This was measured using an online X-ray grating spectrometer \cite{thzx_yang2024_spectrometer}. As shown in Fig.~\ref{fig:3}b, the modulated beam produces equidistant spectral spikes compared to the standard SASE spectrum. The spectral electric field is obtained by Fourier transforming the temporal electric field, $\widetilde{E}(\omega)=\mathcal{F}\{E(t)\}$, and the measured X-ray spectral intensity is proportional to $S(\omega)=\left|\widetilde{E}(\omega)\right|^2$. The  frequency spacing between adjacent spikes is about 19 THz, and the temporal interval ($t_{0j+1}-t_{0j}$) between the peak of the sub-pulses can be calculated as 52.6 fs, which is in good agreement with the intervals between the THz pulse trains in Fig.~\ref{fig:1}b and Fig.~\ref{fig:1}c. Because each sub-pulse builds up from independent shot noise, the phases $\phi_j$ are random, which causes shot-to-shot fluctuations in the exact fringe positions. However, the consistent periodic envelope proves the generation of X-ray pulse trains \cite{thzx_duris2021_pulsetrains,thzx_hu2025_modelockedcomb}.

To evaluate the THz frequency tunability of this scheme, we characterized the system at three representative modulation frequencies ($15$, $19$ and $30\text{ THz}$). Fig.~\ref{fig:4}a shows the measured THz pulse energies and spectra (obtained using a Michelson interferometer). The THz emission reaches a maximum pulse energy of $595\ \mu\text{J}$ at $30\text{ THz}$, with narrow relative spectral bandwidths ranging from $8.6\%$ to $10.4\%$. These measurements show that the THz pulse energy and spectral bandwidth remain basically unchanged whether the upstream X-ray FEL is lasing or non-lasing. This directly confirms that the intense X-ray radiation has a very limited impact on the downstream THz output. Fig.~\ref{fig:4}b presents the corresponding X-ray pulse energies at these THz frequencies, alongside the unmodulated case for comparison. These results show that the THz modulation consistently enhances the X-ray radiation. For example, the mean X-ray pulse energy increases from $122\ \mu\text{J}$ to $156\ \mu\text{J}$ with the $15\text{ THz}$ modulation, and from $133\ \mu\text{J}$ to $225\ \mu\text{J}$ with the $30\text{ THz}$ modulation. This enhancement is well supported by the simulated results. Finally, transverse profile measurements (Fig.~\ref{fig:1}d, e) taken with an X-ray fluorescent screen and a THz camera (Dolphin Optics) show that both the X-ray ($1.2 \times 1.2\text{ mm}^2$ FWHM) and focused THz ($145 \times 138\ \mu\text{m}^2$ FWHM) beams maintain regular spatial modes. These results demonstrate that the THz modulation can enhance the X-ray pulse energy while preserving the overall radiation quality.

In summary, we have experimentally demonstrated the generation of intrinsically synchronized THz and X-ray FELs from a single pre-bunched electron beam. The results show that the THz modulation boosts the X-ray pulse energy without degrading the spectral bandwidth. The initial THz bunching survives the intense X-ray lasing process, allowing the downstream generation of high-power THz radiation. In our experiment, we simultaneously produced tunable THz pulses ($15$ to $30\text{ THz}$, up to $595\ \mu\text{J}$) and enhanced X-ray pulses (up to $225\ \mu\text{J}$ at $4.1\text{ nm}$). This single-beam setup simplifies the experimental layout and preserves the benefits of the electron-beam tailoring scheme \cite{thzx_kang2026_continuous,thzx_kang2023_xfelthz}, such as high pulse energy, wide tunability, and narrow bandwidth. Furthermore, the pre-bunched electron beam shapes the X-ray FEL into pulse trains, which offer a route to probe electronic dynamics with atomic-site specificity and provide opportunities for spectral manipulation. Because each X-ray sub-pulse is locked in time to the same THz waveform that drives the downstream radiation, the pulse train could, in principle, sample several phase points of a single THz pulse within one shot. This offers a route to single-shot, multi-time-point probing of THz-driven ultrafast dynamics, which is of particular value for irreversible or poorly reproducible processes.

The spatial separation of the X-ray and THz beams also allows them to serve independent endstations, directly improving the bunch utilization efficiency of the facility. The proposed mechanism is independent of the beam repetition rate, making it fully compatible with high-repetition-rate FEL operations. Operating at lower THz frequencies requires a deeper initial energy modulation, which may reduce the X-ray power; however, this trade-off can be managed by optimizing the bunch compression settings. Due to the sequential layout, the X-ray pulse naturally precedes the THz pulse. This fixed time ordering is immediately applicable to X-ray-pump/THz-probe experiments. For the THz-pump/X-ray-probe studies, the temporal sequence can be adjusted using X-ray multilayer delay lines \cite{thzx_sauppe2018_splitdelay}, which are well suited for our fixed-wavelength X-ray operation. Overall, this jitter-free, dual-wavelength source establishes a highly reliable platform for advanced ultrafast pump-probe spectroscopy.

\section{ACKNOWLEDGMENTS}

The authors would like to thank Enrico Allaria for helpful discussions. This work was supported by the National Natural Science Foundation of China under grant no.~12275340, 12105347 and 12435011, CAS Project for Young Scientists in Basic Research under grant no.~YSBR-115, China Postdoctoral Science Foundation under grant no.~BX2026146, Shanghai Municipal Science and Technology Major Project and Innovation Program of Shanghai Advanced Research Institute, CAS under grant no.~2024CP001.

\section{Data availability}
The data are available from the authors upon reasonable request.

\bibliography{sxfel_thz_xray_refs}

@article{thzx_basini2024_multiferroicity,
  title={Terahertz electric-field-driven dynamical multiferroicity in {SrTiO$_3$}},
  author={Basini, Martina and Pancaldi, Matteo and Wehinger, Bj{\"o}rn and Udina, Mattia and Unikandanunni, V. and Tadano, T. and Hoffmann, M. C. and Balatsky, A. V. and Bonetti, Stefano},
  journal={Nature},
  volume={628},
  pages={534--539},
  year={2024},
  publisher={Springer Nature},
  doi={10.1038/s41586-024-07175-9}
}

@article{thzx_fechner2024_lattice,
  title={Quenched lattice fluctuations in optically driven {SrTiO$_3$}},
  author={Fechner, Michael and others},
  journal={Nature Materials},
  volume={23},
  pages={363--368},
  year={2024},
  publisher={Springer Nature},
  doi={10.1038/s41563-023-01791-y}
}

@article{thzx_zhang2024_magnon,
  title={Terahertz-field-driven magnon upconversion in an antiferromagnet},
  author={Zhang, Z. and others},
  journal={Nature Physics},
  volume={20},
  pages={788--793},
  year={2024},
  publisher={Springer Nature},
  doi={10.1038/s41567-023-02350-7}
}

@article{thzx_ilyas2024_magnetization,
  title={Terahertz field-induced metastable magnetization near criticality in {FePS$_3$}},
  author={Ilyas, Batyr and others},
  journal={Nature},
  volume={636},
  number={8043},
  pages={609--614},
  year={2024},
  publisher={Springer Nature},
  doi={10.1038/s41586-024-08226-x}
}

@article{thzx_mashkovich2021_spinlattice,
  title={Terahertz light--driven coupling of antiferromagnetic spins to lattice},
  author={Mashkovich, Evgeny A. and Grishunin, Kirill A. and Dubrovin, Roman M. and Zvezdin, Anatoly K. and Pisarev, Roman V. and Kimel, Alexey V.},
  journal={Science},
  volume={374},
  number={6575},
  pages={1608--1611},
  year={2021},
  publisher={American Association for the Advancement of Science},
  doi={10.1126/science.abk1121}
}

@article{thzx_yamakawa2021_chargeorder,
  title={Terahertz-field-induced polar charge order in electronic-type dielectrics},
  author={Yamakawa, Hiroshi and Miyamoto, Tatsuya and Morimoto, Takahiro and Takamura, N. and Liang, S. and Yoshimochi, H. and Terashige, T. and Kida, N. and Suda, M. and Yamamoto, H. M. and Mori, H. and Miyagawa, K. and Kanoda, K. and Okamoto, H.},
  journal={Nature Communications},
  volume={12},
  number={1},
  pages={953},
  year={2021},
  publisher={Springer Nature},
  doi={10.1038/s41467-021-20925-x}
}

@article{thzx_shi2023_mote2,
  title={Intrinsic {1T$^{\prime}$} phase induced in atomically thin {2H-MoTe$_2$} by a single terahertz pulse},
  author={Shi, Jiaojian and Bie, Ya-Qing and Zong, Alfred and Fang, Shiang and Chen, Wei and Han, Jinchi and Cao, Zhaolong and Zhang, Yong and Taniguchi, Takashi and Watanabe, Kenji and Fu, Xuewen and Bulovic, Vladimir and Kaxiras, Efthimios and Baldini, Edoardo and Jarillo-Herrero, Pablo and Nelson, Keith A.},
  journal={Nature Communications},
  volume={14},
  number={1},
  pages={5905},
  year={2023},
  publisher={Springer Nature},
  doi={10.1038/s41467-023-41291-w}
}

@article{thzx_ueda2023_spinlattice,
  title={Non-equilibrium dynamics of spin-lattice coupling},
  author={Ueda, Hiroki and Mankowsky, Roman and Paris, Eugenio and others},
  journal={Nature Communications},
  volume={14},
  number={1},
  pages={7778},
  year={2023},
  publisher={Springer Nature},
  doi={10.1038/s41467-023-43581-9}
}

@article{thzx_johnson2023_vo2,
  title={Ultrafast {X-ray} imaging of the light-induced phase transition in {VO$_2$}},
  author={Johnson, Allan S. and others},
  journal={Nature Physics},
  volume={19},
  pages={215--220},
  year={2023},
  publisher={Springer Nature},
  doi={10.1038/s41567-022-01848-w}
}

@article{thzx_salen2019_matter_manipulation,
  title={Matter manipulation with extreme terahertz light: Progress in the enabling {THz} technology},
  author={Sal{\'e}n, Peter and Basini, Martina and Bonetti, Stefano and Hebling, J{\'a}nos and Krasilnikov, Mikhail and Nikitin, Alexey Y. and Shamuilov, Georgii and Tibai, Zolt{\'a}n and Zhaunerchyk, Vitali and Goryashko, Vitaliy},
  journal={Physics Reports},
  volume={836--837},
  pages={1--74},
  year={2019},
  publisher={Elsevier},
  doi={10.1016/j.physrep.2019.09.002}
}

@article{thzx_delatorre2021_nonthermal,
  title={Colloquium: Nonthermal pathways to ultrafast control in quantum materials},
  author={de la Torre, Alberto and Kennes, Dante M. and Claassen, Martin and Gerber, Simon and McIver, James W. and Sentef, Michael A.},
  journal={Reviews of Modern Physics},
  volume={93},
  number={4},
  pages={041002},
  year={2021},
  publisher={American Physical Society},
  doi={10.1103/RevModPhys.93.041002}
}

@article{thzx_pellegrini2016_xfelphysics,
  title={The physics of {X-ray} free-electron lasers},
  author={Pellegrini, Claudio and Marinelli, Agostino and Reiche, Sven},
  journal={Reviews of Modern Physics},
  volume={88},
  number={1},
  pages={015006},
  year={2016},
  publisher={American Physical Society},
  doi={10.1103/RevModPhys.88.015006}
}

@article{thzx_kubota2025_xraythz,
  title={A simple method to find temporal overlap between {THz} and {X-ray} pulses using {X-ray}-induced carrier dynamics in semiconductors},
  author={Kubota, Yuya and Suzuki, Takeshi and Owada, Shigeki and Tamasaku, Kenji and Osawa, Hitoshi and Togashi, Tadashi and Okazaki, Kozo and Yabashi, Makina},
  journal={Applied Physics Letters},
  volume={126},
  number={5},
  pages={052101},
  year={2025},
  publisher={AIP Publishing},
  doi={10.1063/5.0242393}
}

@article{thzx_zhang2017_extreme_terahertz_science,
  title={Extreme terahertz science},
  author={Zhang, Xi-Cheng and Shkurinov, Alexander and Zhang, Yan},
  journal={Nature Photonics},
  volume={11},
  number={1},
  pages={16--18},
  year={2017},
  publisher={Springer Nature},
  doi={10.1038/nphoton.2016.249}
}

@article{thzx_kampfrath2013_resonant_nonresonant_control,
  title={Resonant and nonresonant control over matter and light by intense terahertz transients},
  author={Kampfrath, Tobias and Tanaka, Koichiro and Nelson, Keith A.},
  journal={Nature Photonics},
  volume={7},
  number={9},
  pages={680--690},
  year={2013},
  publisher={Springer Nature},
  doi={10.1038/nphoton.2013.184}
}

@article{thzx_zhang2020_lcls2thz,
  title={A high-power, high-repetition-rate {THz} source for pump--probe experiments at {Linac Coherent Light Source II}},
  author={Zhang, Z. and Fisher, A. S. and Hoffmann, M. C. and Jacobson, B. and Kirchmann, P. S. and others},
  journal={Journal of Synchrotron Radiation},
  volume={27},
  number={4},
  pages={890--901},
  year={2020},
  publisher={International Union of Crystallography},
  doi={10.1107/S1600577520005147}
}

@article{thzx_kang2026_continuous,
  title={Continuous terahertz band coverage through precise electron-beam tailoring in free-electron lasers},
  author={Kang, Yin and Li, Tong and Wang, Zhen and Wang, Yue and Yu, Cheng and Yin, Weiyi and Gao, Zhangfeng and Xu, Hanghua and Luo, Hang and Wang, Xiaofan and others},
  journal={Nature Photonics},
  volume={20},
  number={1},
  pages={96--101},
  year={2026},
  publisher={Springer Nature},
  doi={10.1038/s41566-025-01775-1}
}

@article{thzx_kang2023_xfelthz,
  title={Generating high-power, frequency tunable coherent {THz} pulse in an {X-ray} free-electron laser for {THz} pump and {X-ray} probe experiments},
  author={Kang, Yin and Wang, Zhen and Zhang, Kaiqing and Feng, Chao},
  journal={Photonics},
  volume={10},
  number={2},
  pages={133},
  year={2023},
  publisher={MDPI},
  doi={10.3390/photonics10020133}
}

@article{thzx_wu2013_ctr,
  title={Intense terahertz pulses from {SLAC} electron beams using coherent transition radiation},
  author={Wu, Z. and Fisher, A. S. and Goodfellow, J. and Fuchs, M. and Daranciang, D. and Hogan, M. and Loos, H. and Lindenberg, A. M.},
  journal={Review of Scientific Instruments},
  volume={84},
  number={2},
  pages={022701},
  year={2013},
  publisher={AIP Publishing},
  doi={10.1063/1.4790427}
}

@article{thzx_weling1996_narrowband,
  title={Novel sources and detectors for coherent tunable narrow-band terahertz radiation in free space},
  author={Weling, A. S. and Auston, D. H.},
  journal={Journal of the Optical Society of America B},
  volume={13},
  number={12},
  pages={2783--2792},
  year={1996},
  publisher={Optica Publishing Group},
  doi={10.1364/JOSAB.13.002783}
}

@article{thzx_dunning2012_periodic,
  title={Generating periodic terahertz structures in a relativistic electron beam through frequency down-conversion of optical lasers},
  author={Dunning, M. and Hast, C. and Hemsing, E. and Jobe, K. and McCormick, D. and Nelson, J. and Raubenheimer, T. O. and Soong, K. and Szalata, Z. and Walz, D. and Weathersby, S. and Xiang, D.},
  journal={Physical Review Letters},
  volume={109},
  number={7},
  pages={074801},
  year={2012},
  publisher={American Physical Society},
  doi={10.1103/PhysRevLett.109.074801}
}

@article{thzx_musumeci2011_spacecharge,
  title={Nonlinear longitudinal space charge oscillations in relativistic electron beams},
  author={Musumeci, P. and Li, R. K. and Marinelli, A.},
  journal={Physical Review Letters},
  volume={106},
  number={18},
  pages={184801},
  year={2011},
  publisher={American Physical Society},
  doi={10.1103/PhysRevLett.106.184801}
}

@article{thzx_saldin2004_microbunching,
  title={Longitudinal space charge-driven microbunching instability in the {TESLA} Test Facility linac},
  author={Saldin, E. L. and Schneidmiller, E. A. and Yurkov, M. V.},
  journal={Nuclear Instruments and Methods in Physics Research Section A: Accelerators, Spectrometers, Detectors and Associated Equipment},
  volume={528},
  number={1},
  pages={355--359},
  year={2004},
  publisher={Elsevier},
  doi={10.1016/j.nima.2004.04.067}
}

@article{thzx_bonifacio1984_highgain,
  title={Collective instabilities and high-gain regime in a free electron laser},
  author={Bonifacio, R. and Pellegrini, C. and Narducci, L. M.},
  journal={Optics Communications},
  volume={50},
  number={6},
  pages={373--378},
  year={1984},
  publisher={Elsevier},
  doi={10.1016/0030-4018(84)90105-6}
}

@book{thzx_saldin2000_felbook,
  title={The Physics of Free Electron Lasers},
  author={Saldin, E. L. and Schneidmiller, E. A. and Yurkov, M. V.},
  year={2000},
  publisher={Springer},
  address={Berlin}
}

@inproceedings{thzx_xie1995_optimization,
  title={Design optimization for an {X-ray} free electron laser driven by {SLAC} linac},
  author={Xie, Ming},
  booktitle={Proceedings of the 1995 Particle Accelerator Conference},
  volume={1},
  pages={183--185},
  year={1995},
  publisher={IEEE},
  address={Dallas, TX, USA},
  doi={10.1109/PAC.1995.504603}
}

@article{thzx_zholents2005_esase,
  title={Method of an enhanced self-amplified spontaneous emission for {X-ray} free electron lasers},
  author={Zholents, Alexander A.},
  journal={Physical Review Special Topics--Accelerators and Beams},
  volume={8},
  number={4},
  pages={040701},
  year={2005},
  publisher={American Physical Society},
  doi={10.1103/PhysRevSTAB.8.040701}
}

@manual{thzx_flottmann2017_astra,
  title={{{ASTRA}: A Space Charge Tracking Algorithm}},
  author={Fl{\"o}ttmann, Klaus},
  year={2017},
  organization={DESY},
  address={Hamburg, Germany},
  url={https://www.desy.de/~mpyflo/Astra_manual/Astra-Manual_V3.2.pdf}
}

@techreport{thzx_borland2000_elegant,
  title={{{ELEGANT}: A Flexible {SDDS}-Compliant Code for Accelerator Simulation}},
  author={Borland, Michael},
  number={LS-287},
  year={2000},
  institution={Advanced Photon Source, Argonne National Laboratory},
  month={Sep},
  doi={10.2172/761286}
}

@article{thzx_reiche1999_genesis,
  title={{{GENESIS} 1.3: A fully 3D time-dependent {FEL} simulation code}},
  author={Reiche, Sven},
  journal={Nuclear Instruments and Methods in Physics Research Section A: Accelerators, Spectrometers, Detectors and Associated Equipment},
  volume={429},
  number={1--3},
  pages={243--248},
  year={1999},
  publisher={Elsevier},
  doi={10.1016/S0168-9002(99)00114-X}
}

@article{thzx_marinelli2016_opticalshaping,
  title={Optical shaping of {X-ray} free-electron lasers},
  author={Marinelli, A. and Coffee, R. and Vetter, S. and Hering, P. and West, G. N. and Gilevich, S. and Lutman, A. A. and Li, S. and Maxwell, T. J. and Galayda, J. and others},
  journal={Physical Review Letters},
  volume={116},
  number={25},
  pages={254801},
  year={2016},
  publisher={American Physical Society},
  doi={10.1103/PhysRevLett.116.254801}
}

@article{thzx_duris2021_pulsetrains,
  title={Controllable {X-ray} pulse trains from enhanced self-amplified spontaneous emission},
  author={Duris, Joseph P. and MacArthur, James P. and Glownia, James M. and Li, Siqi and Vetter, Sharon and Miahnahri, Alan and Coffee, Ryan and Hering, Philippe and Fry, Alan and Welch, Marc E. and others},
  journal={Physical Review Letters},
  volume={126},
  number={10},
  pages={104802},
  year={2021},
  publisher={American Physical Society},
  doi={10.1103/PhysRevLett.126.104802}
}

@article{thzx_hu2025_modelockedcomb,
  title={Demonstration of Mode-Locked Frequency Comb for an {X-Ray} Free-Electron Laser},
  author={Hu, Wenxiang and Aeppli, Gabriel and Arrell, Christopher and Calvi, Marco and Carbajo, Sergio and Dax, Andreas and Deng, Yunpei and Dijkstal, Philipp and Dunning, David and Gerber, Simon and Huppert, Martin and Neppl, Stefan and Reiche, Sven and Schietinger, Thomas and Thompson, Neil and Trisorio, Alexandre and Vicario, Carlo and Zholents, Alexander and Prat, Eduard},
  journal={Physical Review Letters},
  volume={135},
  number={26},
  pages={265001},
  year={2025},
  publisher={American Physical Society},
  doi={10.1103/wn8d-l7sh}
}

@article{thzx_yang2024_spectrometer,
  title={Development and commissioning of a broadband online {X-ray} spectrometer for the {SXFEL} Facility},
  author={Yang, Zhicheng and Zhang, Ximing and Geng, Heping and Chen, Jiahua and Feng, Chao and Liu, Bo and Li, Bin},
  journal={Journal of Synchrotron Radiation},
  volume={31},
  number={5},
  pages={1373--1381},
  year={2024},
  publisher={International Union of Crystallography},
  doi={10.1107/S1600577524005812}
}

@article{thzx_sauppe2018_splitdelay,
  title={{{XUV} double-pulses with femtosecond to 650 ps separation from a multilayer-mirror-based split-and-delay unit at {FLASH}}},
  author={Sauppe, Mario and Rompotis, Dimitrios and Erk, Benjamin and Bari, Sadia and Bischoff, Tobias and Boll, Rebecca and Bomme, Cedric and Bostedt, Christoph and D{\"o}rner, Simon and D{\"u}sterer, Stefan and others},
  journal={Journal of Synchrotron Radiation},
  volume={25},
  number={5},
  pages={1517--1528},
  year={2018},
  publisher={International Union of Crystallography},
  doi={10.1107/S1600577518006094}
}

@article{thzx_johnson2025_xrayview,
  title={Perspective: an {X-ray} view of the coherent driving of materials},
  author={Johnson, Steven L. and Staub, Urs},
  journal={npj Quantum Materials},
  volume={10},
  pages={78},
  year={2025},
  publisher={Springer Nature},
  doi={10.1038/s41535-025-00799-8}
}

@article{thzx_li2019_ferroelectricity,
  title={Terahertz field-induced ferroelectricity in quantum paraelectric {SrTiO$_3$}},
  author={Li, Xian and Qiu, Tian and Zhang, Jiahao and Baldini, Edoardo and Lu, Jian and Rappe, Andrew M. and Nelson, Keith A.},
  journal={Science},
  volume={364},
  number={6445},
  pages={1079--1082},
  year={2019},
  publisher={American Association for the Advancement of Science},
  doi={10.1126/science.aaw4913}
}

@article{thzx_kubacka2014_electromagnon,
  title={Large-amplitude spin dynamics driven by a {THz} pulse in resonance with an electromagnon},
  author={Kubacka, T. and Johnson, J. A. and Hoffmann, M. C. and Vicario, C. and de Jong, S. and Beaud, P. and Gr{\"u}bel, S. and Huang, S.-W. and Huber, L. and Patthey, L. and Chuang, Y.-D. and Turner, J. J. and Dakovski, G. L. and Lee, W.-S. and Minitti, M. P. and Schlotter, W. and Moore, R. G. and Hauri, C. P. and Koohpayeh, S. M. and Scagnoli, V. and Ingold, G. and Johnson, S. L. and Staub, U.},
  journal={Science},
  volume={343},
  number={6177},
  pages={1333--1336},
  year={2014},
  publisher={American Association for the Advancement of Science},
  doi={10.1126/science.1242862}
}

@article{thzx_li2021_polarvortices,
  title={Subterahertz collective dynamics of polar vortices},
  author={Li, Qian and Stoica, Vladimir A. and Pa{\'s}ciak, Marek and Zhu, Yi and Yuan, Yakun and Yang, Tiannan and McCarter, Margaret R. and Das, Sujit and Yadav, Ajay K. and Park, Suji and Dai, Cheng and Lee, Hyeon Jun and Ahn, Youngjun and Marks, Samuel D. and Yu, Shukai and Kadlec, Christelle and Sato, Takahiro and Hoffmann, Matthias C. and Chollet, Matthieu and Kozina, Michael E. and Nelson, Silke and Zhu, Diling and Walko, Donald A. and Lindenberg, Aaron M. and Evans, Paul G. and Chen, Long-Qing and Ramesh, Ramamoorthy and Martin, Lane W. and Gopalan, Venkatraman and Freeland, John W. and Hlinka, Jirka and Wen, Haidan},
  journal={Nature},
  volume={592},
  pages={376--380},
  year={2021},
  publisher={Springer Nature},
  doi={10.1038/s41586-021-03342-4}
}

@article{thzx_orenstein2025_polarizationwaves,
  title={Observation of polarization density waves in {SrTiO$_3$}},
  author={Orenstein, Gal and Krapivin, Viktor and Huang, Yijing and Zhang, Zhuquan and de la Pe{\~n}a Mu{\~n}oz, Gilberto and Duncan, Ryan A. and Nguyen, Quynh and Stanton, Jade and Teitelbaum, Samuel and Yavas, Hasan and Sato, Takahiro and Hoffmann, Matthias C. and Kramer, Patrick and Zhang, Jiahao and Cavalleri, Andrea and Comin, Riccardo and Dean, Mark P. M. and Disa, Ankit S. and F{\"o}rst, Michael and Johnson, Steven L. and Mitrano, Matteo and Rappe, Andrew M. and Reis, David A. and Zhu, Diling and Nelson, Keith A. and Trigo, Mariano},
  journal={Nature Physics},
  volume={21},
  pages={961--965},
  year={2025},
  publisher={Springer Nature},
  doi={10.1038/s41567-025-02874-0}
}

@article{thzx_venanzi2024_trions,
  title={Ultrafast switching of trions in {2D} materials by terahertz photons},
  author={Venanzi, Tommaso and Cuccu, Marzia and Perea-Causin, Raul and Sun, Xiaoxiao and Brem, Samuel and Erkensten, Daniel and Taniguchi, Takashi and Watanabe, Kenji and Malic, Ermin and Helm, Manfred and Winnerl, Stephan and Chernikov, Alexey},
  journal={Nature Photonics},
  volume={18},
  pages={1344--1349},
  year={2024},
  publisher={Springer Nature},
  doi={10.1038/s41566-024-01512-0}
}

@article{thzx_zapolnova2020_plasmaswitch,
  title={{XUV}-driven plasma switch for {THz}: new spatio-temporal overlap tool for {XUV--THz} pump--probe experiments at {FELs}},
  author={Zapolnova, E. and Pan, R. and Golz, T. and Sindik, M. and Nikolic, M. and Temme, M. and Rabasovic, M. and Grujic, D. and Chen, Z. and Toleikis, S. and Stojanovic, N.},
  journal={Journal of Synchrotron Radiation},
  volume={27},
  number={1},
  pages={11--16},
  year={2020},
  publisher={International Union of Crystallography},
  doi={10.1107/S1600577519014164}
}

@article{thzx_chen2021_ultrafast_conductivity,
  title={Ultrafast multi-cycle terahertz measurements of the electrical conductivity in strongly excited solids},
  author={Chen, Z. and Curry, C. B. and Zhang, R. and Treffert, F. and Stojanovic, N. and Toleikis, S. and Pan, R. and Gauthier, M. and Zapolnova, E. and Seipp, L. E. and Weinmann, A. and Mo, M. Z. and Kim, J. B. and Witte, B. B. L. and Bajt, S. and Usenko, S. and Soufli, R. and Pardini, T. and Hau-Riege, S. and Burcklen, C. and Schein, J. and Redmer, R. and Tsui, Y. Y. and Ofori-Okai, B. K. and Glenzer, S. H.},
  journal={Nature Communications},
  volume={12},
  number={1},
  pages={1638},
  year={2021},
  publisher={Springer Nature},
  doi={10.1038/s41467-021-21756-6}
}

@article{thzx_kozina2019_upconversion,
  author    = {Kozina, Michael
               and Fechner, Michael
               and Marsik, Premysl
               and van Driel, Tim
               and Glownia, James M.
               and Bernhard, Christian
               and Radovic, Milan
               and Zhu, Diling
               and Bonetti, Stefano
               and Staub, Urs
               and Hoffmann, Matthias C.},
  title     = {Terahertz-driven phonon upconversion in {SrTiO$_3$}},
  journal   = {Nature Physics},
  volume    = {15},
  number    = {4},
  pages     = {387--392},
  year      = {2019},
  publisher = {Springer Nature},
  doi       = {10.1038/s41567-018-0408-1}
}

@article{thzx_wang2025_skyrons,
  author    = {Wang, Huaiyu Hugo
               and Stoica, Vladimir A.
               and Dai, Cheng
               and Pa{\'s}ciak, Marek
               and Das, Sujit
               and Yang, Tiannan
               and Gon{\c{c}}alves, Mauro A. P.
               and Kulda, Jiri
               and McCarter, Margaret R.
               and Mangu, Anudeep
               and Cao, Yue
               and Padma, Hari
               and Saha, Utkarsh
               and Zhu, Diling
               and Sato, Takahiro
               and Song, Sanghoon
               and Hoffmann, Matthias C.
               and Kramer, Patrick
               and Nelson, Silke
               and Sun, Yanwen
               and Nguyen, Quynh
               and Zhang, Zhan
               and Ramesh, Ramamoorthy
               and Martin, Lane W.
               and Lindenberg, Aaron M.
               and Chen, Long Qing
               and Freeland, John W.
               and Hlinka, Jirka
               and Gopalan, Venkatraman
               and Wen, Haidan},
  title     = {Terahertz-field activation of polar skyrons},
  journal   = {Nature Communications},
  volume    = {16},
  number    = {1},
  pages     = {8994},
  year      = {2025},
  publisher = {Springer Nature},
  doi       = {10.1038/s41467-025-64033-6}
}

@article{thzx_pan2019_flash_diagnostics,
  author = {Pan, Rui
            and Zapolnova, Ekaterina
            and Golz, Torsten
            and Krmpot, Aleksandar J.
            and Rabasovic, Mihailo D.
            and Petrovic, Jovana
            and Asgekar, Vivek
            and Faatz, Bart
            and Tavella, Franz
            and Perucchi, Andrea
            and Kovalev, Sergey
            and Green, Bertram
            and Geloni, Gianluca
            and Tanikawa, Takanori
            and Yurkov, Mikhail
            and Schneidmiller, Evgeny
            and Gensch, Michael
            and Stojanovic, Nikola},
  title = {Photon diagnostics at the {FLASH} {THz} beamline},
  journal = {Journal of Synchrotron Radiation},
  volume = {26},
  number = {3},
  pages = {700--707},
  year = {2019},
  doi = {10.1107/S1600577519003412},
  publisher = {International Union of Crystallography}
}

@article{thzx_zapolnova2018_flash_doubler,
  author = {Zapolnova, Ekaterina
            and Golz, Torsten
            and Pan, Rui
            and Klose, Karsten
            and Schreiber, Siegfried
            and Stojanovic, Nikola},
  title = {{THz} pulse doubler at {FLASH}: double pulses for
           pump--probe experiments at {X-ray FELs}},
  journal = {Journal of Synchrotron Radiation},
  volume = {25},
  number = {1},
  pages = {39--43},
  year = {2018},
  doi = {10.1107/S1600577517015442},
  publisher = {International Union of Crystallography}
}

@article{thzx_dimitri2018_terafermi_ctr,
  author = {Di Mitri, S.
            and Perucchi, A.
            and Adhlakha, N.
            and Di Pietro, P.
            and Nicastro, S.
            and Roussel, E.
            and Spampinati, S.
            and Veronese, M.
            and Allaria, E.
            and Badano, L.
            and Cudin, I.
            and De Ninno, G.
            and Diviacco, B.
            and Gaio, G.
            and Gauthier, D.
            and Giannessi, L.
            and Lupi, S.
            and Penco, G.
            and Piccirilli, F.
            and Rebernik, P.
            and Spezzani, C.
            and Trov{\`o}, M.},
  title = {Coherent {THz} Emission Enhanced by Coherent
           Synchrotron Radiation Wakefield},
  journal = {Scientific Reports},
  volume = {8},
  number = {1},
  pages = {11661},
  year = {2018},
  publisher = {Springer Nature},
  doi = {10.1038/s41598-018-30125-1}
}

@article{thzx_brynes2020_mbi_modulated_laser,
  author = {Brynes, A. D.
            and Akkermans, I.
            and Allaria, E.
            and Badano, L.
            and Brussaard, S.
            and Danailov, M.
            and Demidovich, A.
            and De Ninno, G.
            and Giannessi, L.
            and Mirian, N. S.
            and Penco, G.
            and Perosa, G.
            and Rebernik Ribi{\v{c}}, P.
            and Roussel, E.
            and Setija, I.
            and Smorenburg, P.
            and Spampinati, S.
            and Spezzani, C.
            and Trov{\`o}, M.
            and Williams, P. H.
            and Wolski, A.
            and Di Mitri, S.},
  title = {Microbunching instability characterization via
           temporally modulated laser pulses},
  journal = {Physical Review Accelerators and Beams},
  volume = {23},
  number = {10},
  pages = {104401},
  year = {2020},
  publisher = {American Physical Society},
  doi = {10.1103/PhysRevAccelBeams.23.104401}
}

@article{thzx_guo2026_plasmonic_time_crystal,
  author = {Guo, Tingwen
            and Sueiro, Jules
            and Andolina, Gian Marcello
            and Levchuk, Artem
            and Ponzoni, Stefano
            and Grasset, Romain
            and Monthe, Donald
            and Aupiais, Ian
            and Daineka, Dmitri
            and Briatico, Javier
            and de Oliveira, Thales V. A. G.
            and Ponomaryov, Alexey
            and Arshad, Atiqa
            and Karimbana-Kandy, Arjun
            and Prajapati, Gulloo Lal
            and Ilyakov, Igor
            and Deinert, Jan-Christoph
            and Maehrlein, Sebastian F.
            and Perfetti, Luca
            and Schir{\`o}, Marco
            and Laplace, Yannis},
  title = {Plasmonic metamaterial time crystal},
  journal = {Nature},
  year = {2026},
  month = jul,
  doi = {10.1038/s41586-026-10825-9},
  note = {Published online 29 July 2026},
  publisher = {Springer Nature}
}

@article{thzx_riepp2024_coherent_magnetization,
  author = {Riepp, M.
            and Philippi-Kobs, A.
            and M{\"u}ller, L.
            and Roseker, W.
            and Rysov, R.
            and Fr{\"o}mter, R.
            and Bagschik, K.
            and Hennes, M.
            and Gupta, D.
            and Marotzke, S.
            and Walther, M.
            and Bajt, S.
            and Pan, R.
            and Golz, T.
            and Stojanovic, N.
            and Boeglin, C.
            and Gr{\"u}bel, G.},
  title = {Terahertz-driven coherent magnetization dynamics
           in labyrinth-type domain networks},
  journal = {Physical Review B},
  volume = {110},
  number = {9},
  pages = {094405},
  year = {2024},
  month = sep,
  publisher = {American Physical Society},
  doi = {10.1103/PhysRevB.110.094405}
}
\bibliographystyle{apsrev4-2-first3-et-al}
\end{document}